\documentclass[10pt,leqno]{amsart}
\usepackage{graphicx}
\usepackage{listings}
\usepackage{indentfirst,csquotes}
\usepackage{siunitx}
\usepackage{booktabs} % For better tables

\lstdefinestyle{mystyle}{
    language=C++,
    backgroundcolor=\color{black!5}, % light gray background
    commentstyle=\color{green!40!black},
    keywordstyle=\color{blue},
    numberstyle=\tiny\color{black!50},
    stringstyle=\color{purple},
    basicstyle=\footnotesize\ttfamily, % smaller, fixed-width font
    breakatwhitespace=false,         
    breaklines=true,                 
    captionpos=b,                    
    keepspaces=true,                 
    numbers=left,                    
    numbersep=5pt,                  
    showspaces=false,                
    showstringspaces=false,
    showtabs=false,                  
    tabsize=2,
    frame=single, % adds a frame around the code
    rulecolor=\color{black}
}

\usepackage{amssymb,amsthm,amsmath}
\usepackage{xcolor,paralist,hyperref,fancyhdr,etoolbox}

\makeatletter
\def\section{\@startsection{section}{1}%
  \z@{.7\linespacing\@plus\linespacing}{.5\linespacing}%
  {\normalfont\Large\bfseries\centering}}
\makeatother

\hypersetup{ colorlinks=true, linkcolor=black, filecolor=black, urlcolor=black }

\begin{document}

\title[Accelerating Matrix Multiplication]{Accelerating Matrix Multiplication: A Performance Comparison Between Multi-Core CPU and GPU}
% --- START: AUTHOR INFORMATION BLOCK ---

\author[M. Q. Ansari]{Mufakir Qamar Ansari}
\address{Department of Electrical Engineering and Computer Science, The University of Toledo, Toledo, Ohio 43606, USA}
\email{mufakir.ansari@utoledo.edu}

\author[M. Q. Ansari]{Mudabir Qamar Ansari}
\address{Department of School of Accounting and Information Systems, Lamar University, Beaumont, Texas 77710, USA}
\email{mansari2@lamar.edu}

\date{\today}

% --- END: AUTHOR INFORMATION BLOCK ---
\subjclass[2020]{Primary 68W10; Secondary 65Y05, 68M20}

\keywords{High-Performance Computing, GPU, CUDA, OpenMP, Matrix Multiplication, Parallel Computing}

\begin{abstract}
Matrix multiplication is a foundational operation in scientific computing and machine learning, yet its computational complexity makes it a significant bottleneck for large-scale applications. The shift to parallel architectures, primarily multi-core CPUs and many-core GPUs, is the established solution, and these systems are now ubiquitous from datacenters to consumer laptops. This paper presents a direct, empirical performance analysis of matrix multiplication on a modern, consumer-grade heterogeneous platform. We implemented and benchmarked three versions of the algorithm: a baseline sequential C++ implementation, a parallel version for its multi-core CPU using OpenMP, and a massively parallel version for its discrete GPU using CUDA with shared memory optimizations. The implementations were evaluated with square matrices of varying dimensions, from 128x128 to 4096x4096. Our results show that while the parallel CPU provides a consistent speedup of 12-14x over the sequential version, the GPU's performance scales dramatically with problem size. For a 4096x4096 matrix, the GPU implementation achieved a speedup of approximately 593x over the sequential baseline and 45x over the optimized parallel CPU version. These findings quantitatively demonstrate the profound impact of many-core GPU architectures on accelerating data-parallel workloads, underscoring that significant performance gains are readily accessible even on consumer-level hardware.
\end{abstract}

\maketitle

\section{Introduction}

\subsection{Context and Motivation}

Matrix-matrix multiplication constitutes a foundational computational kernel, underpinning a vast spectrum of algorithms in scientific computing, data science, and artificial intelligence \cite{eijkhout2010introduction}. Its applications are pervasive, forming the computational core of linear algebra libraries (BLAS), enabling the training of deep neural networks, and driving simulations in fields from physics to finance \cite{huang2019gpu}. The significance of this operation is further underscored by its central role in emerging paradigms like neuromorphic computing \cite{mohamed2020neuromorphic}. However, the algorithm's asymptotic complexity of O(n³), where n is the matrix dimension, presents a formidable computational challenge. As datasets and model sizes continue to grow exponentially, this operation frequently becomes a primary performance bottleneck, demanding architectural and algorithmic solutions that can deliver extreme-scale performance.

\subsection{The Rise of Parallelism}

The computational demands of such operations have long surpassed the capabilities of sequential processing. The era of improving performance by increasing the clock frequency of single-core processors has concluded, halted by the fundamental physical constraints of power consumption and heat dissipation---a phenomenon often termed the ``power wall'' \cite{asanovic2006landscape}. This architectural inflection point has forced the high-performance computing (HPC) industry to embrace parallelism as the primary path to greater computational power. This has led to the development of two dominant, yet architecturally divergent, classes of processors: the multi-core Central Processing Unit (CPU) and the many-core Graphics Processing Unit (GPU). These parallel architectures are no longer confined to supercomputers; they are now a standard component in commodity hardware, including the consumer-grade laptop used in this study.

CPUs are engineered for low-latency execution on a wide variety of tasks, employing sophisticated control logic and deep cache hierarchies to accelerate single-thread performance. In contrast, GPUs are designed as throughput-oriented engines, featuring thousands of simpler, highly-efficient cores that excel at executing the same operation on massive datasets in parallel \cite{ryoo2008optimization}. The co-location of these distinct architectures in a single system gives rise to the field of heterogeneous computing \cite{mittal2015survey}, where the central challenge is to effectively map computational tasks to the hardware best suited for them.

\subsection{Research Objective}

Against this backdrop of architectural divergence, this paper provides a direct, empirical performance comparison of matrix multiplication on a modern, consumer-grade heterogeneous platform, comprising a multi-core CPU and a many-core GPU. We aim to quantify the performance differences and analyze the scaling behavior of each architecture across a range of problem sizes. To this end, we implement and benchmark a standard sequential C++ version alongside optimized parallel implementations using OpenMP for the CPU \cite{dagum1998openmp} and the CUDA framework for the GPU \cite{nvidiaCUDAguide}. By presenting clear, reproducible data from a widely available hardware configuration, this study seeks to illuminate the practical performance characteristics of these competing hardware paradigms for a workload that is fundamental to the entire HPC ecosystem.

\subsection{Paper Structure}

The remainder of this paper is structured as follows. Section 2 reviews the existing literature on CPU-GPU performance analysis and optimization principles. Section 3 details the methodology, including the specific algorithm implementations and the experimental setup. Section 4 presents the performance results, which are then analyzed and discussed in Section 5. Finally, Section 6 offers our conclusions and suggests directions for future research.
%======================================================================
\section{Related Work}
%======================================================================

The performance characteristics of High-Performance Computing (HPC) architectures have been a subject of intense research, particularly with the rise of General-Purpose computing on Graphics Processing Units (GPGPU). This study builds upon a significant body of work that compares CPU and GPU capabilities, explores optimization principles, and evaluates the programming models used to harness these powerful processors.

\subsection{The Performance Landscape of CPUs and GPUs}

The relative performance of multi-core CPUs and many-core GPUs has been a subject of extensive investigation, often framed by claims of orders-of-magnitude speedups. A seminal study by Lee et al. \cite{lee2010debunking} sought to debunk the "100X GPU vs. CPU myth," demonstrating that when applications are highly optimized for both architectures---leveraging SIMD vectorization and multi-threading on the CPU and CUDA principles on the GPU---the performance gap narrows substantially to an average of 2.5x in favor of the GPU. Our work serves as a contemporary validation of this principle, applying a similar comparative analysis to a foundational HPC kernel on modern hardware.

Subsequent research has consistently shown that the ideal architecture is highly dependent on the algorithm's characteristics. Teodoro et al. \cite{teodoro2015performance} conducted a performance analysis across CPUs, GPUs, and Intel's Many Integrated Core (MIC) architecture, finding that GPUs excel at tasks with regular, predictable data access patterns, a key feature of dense matrix multiplication. This conclusion is echoed by Huang et al. \cite{huang2019gpu}, who analyzed matrix multiplication specifically and noted that the GPU's performance advantage increases dramatically with matrix size---a trend our results confirm. The broader field of heterogeneous computing, which aims to leverage the unique strengths of different processors, is thoroughly surveyed by Mittal and Vetter \cite{mittal2015survey}, who frame the motivation for combining, rather than competing, CPU and GPU resources.

\subsection{Optimization Principles for GPU Computing}

Achieving high performance on many-core GPUs is contingent upon a deep understanding of their underlying architecture; it is not merely a matter of offloading code. An early and influential paper by Fatahalian et al. \cite{fatahalian2004understanding} analyzed the efficiency of GPU algorithms for matrix-matrix multiplication and identified the ratio between arithmetic computation and memory bandwidth as the key limiting factor. This insight remains central to GPU optimization today.

The work by Ryoo et al. \cite{ryoo2008optimization} provides a foundational set of optimization principles for the CUDA programming model. They established that effective management of the GPU's memory hierarchy is paramount for performance. Their work highlights the critical importance of achieving coalesced global memory access to maximize effective bandwidth and of utilizing the on-chip shared memory to reduce latency and increase data reuse. The CUDA kernel implemented in our study directly applies these fundamental principles to minimize data movement and maximize computational throughput. More recent work has explored even more specialized optimizations for non-square or sparse matrices \cite{chen2019tsm2, monakov2010automatically}, demonstrating that performance can be further enhanced by tailoring algorithms to specific data structures.

\subsection{Programming Models and the HPC Ecosystem}

This study utilizes two dominant, industry-standard programming models to harness the parallel capabilities of the target architectures: OpenMP and CUDA. The work by Dagum and Menon \cite{dagum1998openmp} describes OpenMP as a portable, directive-based API designed for accessible, incremental parallelization of code on shared-memory systems, which aligns with our straightforward parallelization of the CPU algorithm. For the GPU, the official NVIDIA CUDA C Programming Guide \cite{nvidiaCUDAguide} serves as the definitive reference for the CUDA execution model, memory hierarchy, and programming interface that our implementation is built upon.

The choice between GPU programming models has also been explored in the literature. Karimi et al. \cite{karimi2010performance} presented a performance comparison between CUDA and OpenCL, the other major GPGPU framework. They found that for NVIDIA hardware, CUDA often delivers superior performance in both data transfer and kernel execution, which they attribute to the tight vertical integration of NVIDIA's hardware, compiler, and API. This finding helps justify our use of CUDA as a representative model for a high-performance GPU implementation. These programming models are essential tools for tackling the challenges and opportunities presented by the industry-wide shift to parallel computing, a landscape defined by Asanovic et al. \cite{asanovic2006landscape}.

%======================================================================
%======================================================================
\section{Methodology}
%======================================================================

To conduct our comparative analysis, we developed and benchmarked three distinct implementations of a square matrix multiplication algorithm ($C = A \times B$). These implementations were designed to represent a baseline sequential case and optimized parallel cases for both multi-core CPU and many-core GPU architectures.

\subsection{Implementations}
\subsubsection{Sequential CPU} The baseline for our study is a standard, "naive" C++ implementation of matrix multiplication. It consists of three nested \texttt{for} loops, iterating through the rows of matrix A, the columns of matrix B, and the inner dimension, respectively. This version serves as the fundamental reference point against which all speedups are calculated.

\subsubsection{Parallel CPU (OpenMP)} To leverage the shared-memory parallelism of the host CPU, the sequential code was annotated with OpenMP directives. Specifically, we used the \texttt{\#pragma omp parallel for collapse(2)} directive applied to the two outer loops. The \texttt{collapse(2)} clause is critical for performance, as it instructs the OpenMP runtime to merge the two nested loops into a single, larger iteration space, allowing for more effective load balancing across the CPU's 16 available threads.

\subsubsection{Parallel GPU (CUDA)} The GPU implementation was developed using the CUDA C++ framework to run on the system's dedicated NVIDIA GPU. A custom kernel was written wherein the workload is partitioned such that each thread is responsible for calculating a single element of the resulting matrix C. To mitigate the high latency of global memory access, which is a primary bottleneck in GPU computing, our kernel makes extensive use of on-chip shared memory. Before computation, threads within a thread block cooperatively load small, contiguous blocks (or tiles) of the input matrices A and B into this fast, shared memory. The subsequent matrix multiplication is then performed using data from this low-latency memory, dramatically reducing global memory traffic and increasing overall computational throughput.

\subsection{Experimental Setup} All benchmarks were executed on a Lenovo IdeaPad Gaming 3 15ACH6 laptop to ensure a consistent hardware and software environment.

\begin{itemize}
    \item \textbf{Hardware:} The test system was configured with the following components:
    \begin{itemize}
        \item \textbf{CPU:} An 8-core, 16-thread AMD Ryzen 7 5800H processor with 16\,MB of L3 cache.
        \item \textbf{GPU:} An NVIDIA GeForce GTX 1650 Mobile GPU with 4096\,MiB of dedicated VRAM. All CUDA computations were executed exclusively on this device.
        \item \textbf{System Memory:} 32\,GB of DDR4 RAM, operating at a configured speed of 2667\,MT/s.
    \end{itemize}
    
    \item \textbf{Software:}
    \begin{itemize}
        \item \textbf{Operating System:} Ubuntu 24.04.2 LTS (Kernel 6.8.0-64-generic).
        \item \textbf{Compilers and Toolchains:} The CPU and GPU codes were compiled using g++ version 12.4.0 and the NVIDIA CUDA Toolkit version 12.2, respectively. Standard \texttt{-O3} optimization flags were enabled for all compilations.
        \item \textbf{GPU Driver:} The system was running NVIDIA Driver version 535.247.01.
    \end{itemize}
    
    \item \textbf{Benchmarking Protocol:} The experiments were conducted on square matrices with dimensions ranging from 128x128 to 4096x4096. Input matrices were populated with random 32-bit floating-point values. The execution time for each implementation was measured using C++'s \texttt{std::chrono::high\_resolution\_clock}. For the GPU implementation, the measured wall-clock time is comprehensive, including the time required for memory allocation on the device, the transfer of input matrices from host memory to device memory (\texttt{cudaMemcpyHostToDevice}), kernel execution, and the transfer of the result matrix back from device memory to host (\texttt{cudaMemcpyDeviceToHost}).
\end{itemize}

\subsection{Performance Metrics}

We used two primary metrics to evaluate and compare the performance of the implementations:
\begin{enumerate}
    \item \textbf{Execution Time:} The total wall-clock time, measured in milliseconds (ms), required to complete the matrix multiplication operation.
    \item \textbf{Speedup:} A dimensionless quantity that quantifies the performance improvement of a parallel implementation relative to the sequential baseline. It is calculated as: $S = T_{\text{sequential}} / T_{\text{parallel}}$.
\end{enumerate}

%======================================================================
\section{Results}
%======================================================================

This section presents the empirical data from our performance benchmarks. We first provide the raw execution times and calculated speedups in a tabular format, followed by a series of visualizations that illustrate the performance trends across the different implementations and matrix sizes.

\subsection{Performance Data}

The execution times for the sequential CPU, parallel CPU (OpenMP), and parallel GPU (CUDA) implementations were recorded across seven different square matrix dimensions, from 128x128 to 4096x4096. From these timings, we calculated the speedup of each parallel approach relative to the sequential baseline, as well as the direct speedup of the GPU over the parallel CPU. The comprehensive results are presented in Table~\ref{tab:results}.

\begin{table}[!ht]
\centering
\caption{Execution Times (in milliseconds) and Speedups for Matrix Multiplication.}
\label{tab:results}
% Resize the table to fit within the column width.
\resizebox{\columnwidth}{!}{%
\begin{tabular}{l S[table-format=6.2] S[table-format=5.2] S[table-format=3.2] c c c}
\toprule
& {\textbf{Seq. CPU}} & {\textbf{Par. CPU}} & {\textbf{Par. GPU}} & \multicolumn{3}{c}{\textbf{Calculated Speedups}} \\
\cmidrule(lr){5-7}
\textbf{Matrix Size (N x N)} & {\textbf{Time (ms)}} & {\textbf{Time (ms)}} & {\textbf{Time (ms)}} & {\textbf{Par. CPU vs Seq.}} & {\textbf{GPU vs Par. CPU}} & {\textbf{GPU vs Seq.}} \\
\midrule
128x128     & 2.18      & 7.10      & 0.26      & 0.31x   & 27.02x  & 8.29x \\
256x256     & 20.70     & 2.89      & 0.40      & 7.16x   & 7.18x   & 51.43x \\
512x512     & 264.37    & 19.43     & 2.10      & 13.60x  & 9.25x   & 125.77x \\
1024x1024   & 3721.52   & 295.86    & 13.35     & 12.58x  & 22.16x  & 278.71x \\
2048x2048   & 44691.46  & 3554.41   & 124.01    & 12.57x  & 28.66x  & 360.38x \\
3072x3072   & 171811.07 & 11998.88  & 332.69    & 14.32x  & 36.07x  & 516.42x \\
4096x4096   & 393280.52 & 30332.07  & 663.24    & 12.97x  & 45.73x  & 592.97x \\
\bottomrule
\end{tabular}
}
\end{table}

\subsection{Performance Visualizations}

To better illustrate the performance characteristics and scaling trends, the data from Table~\ref{tab:results} are visualized in the following figures.

Figure~\ref{fig:exectime} plots the execution time of all three implementations as a function of matrix size. A logarithmic scale is used for the Y-axis (Execution Time) to effectively visualize the data. The execution times span several orders of magnitude, from over 390,000\,ms for the largest sequential run to under 1\,ms for the smallest GPU runs. A linear scale would render the performance differences for the two parallel methods at smaller matrix sizes indistinguishable, while obscuring the overall trend.

\begin{figure}[!ht]
    \centering
    \includegraphics[width=\columnwidth]{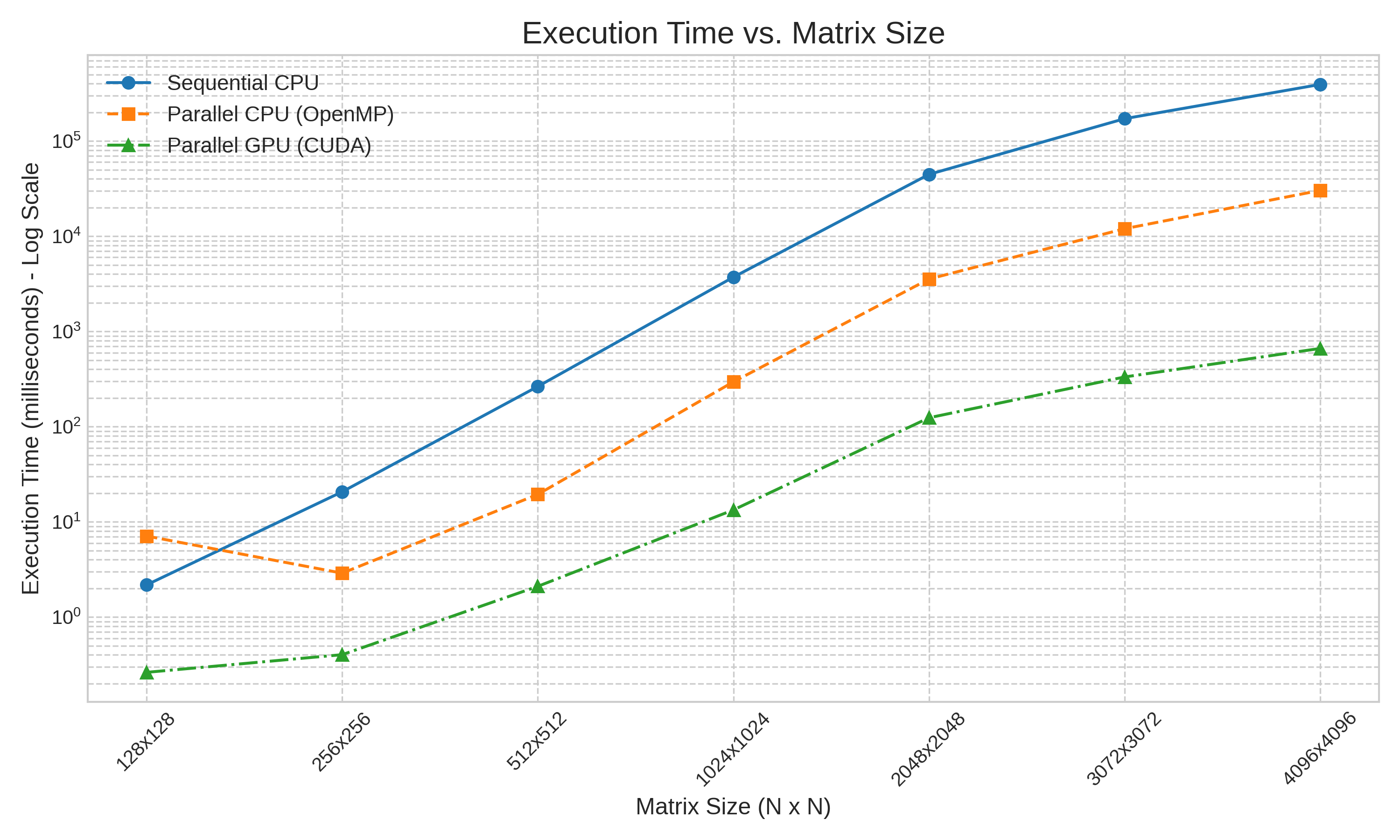}
    \caption{A comparison of execution times for sequential CPU, parallel CPU (OpenMP), and parallel GPU (CUDA) implementations across varying matrix sizes. The logarithmic Y-axis is used to accommodate the wide range of execution times.}
    \label{fig:exectime}
\end{figure}

\begin{figure}[!ht]
    \centering
    \includegraphics[width=\columnwidth]{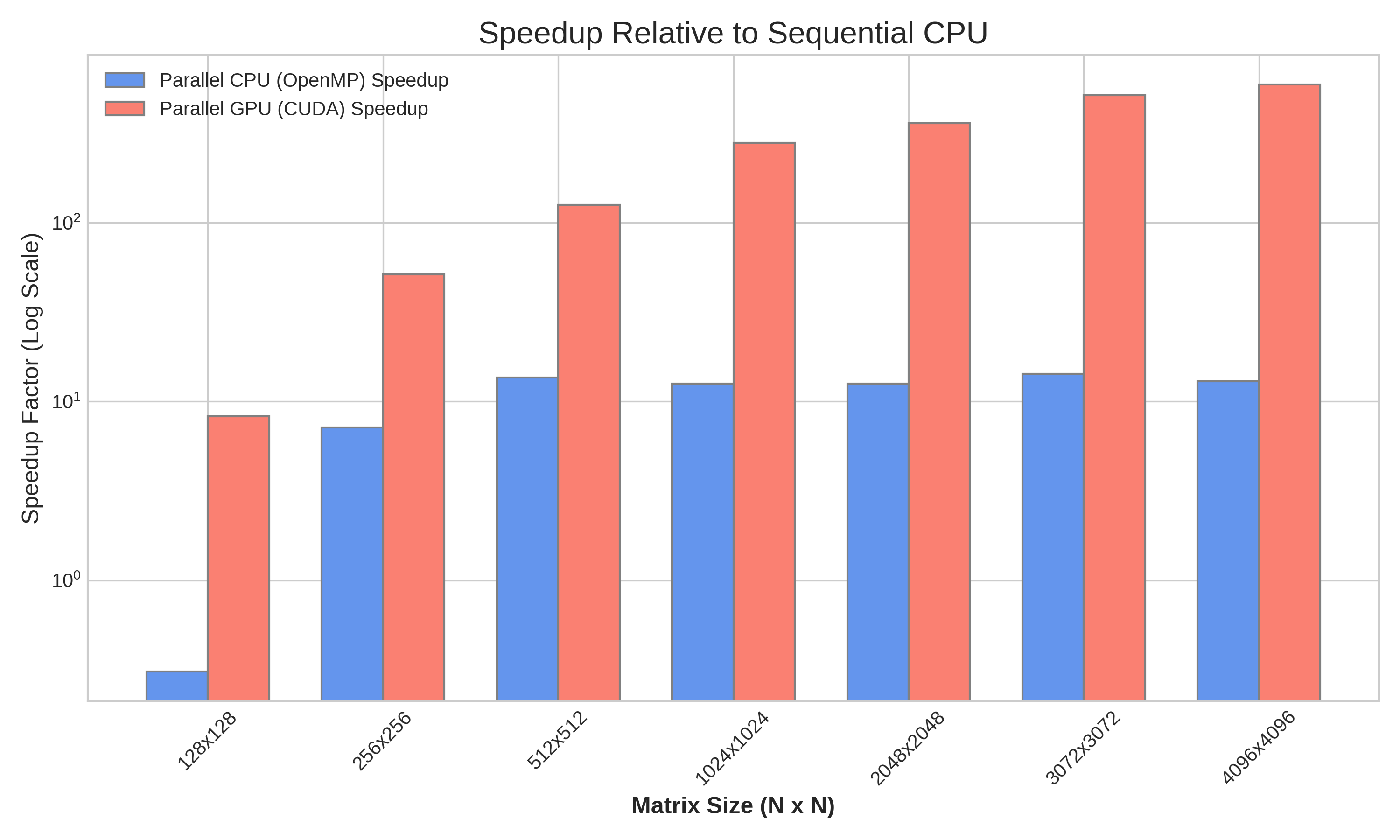}
    \caption{Speedup of the parallel CPU (OpenMP) and parallel GPU (CUDA) implementations relative to the sequential CPU baseline. Note that the Y-axis is on a logarithmic scale to clearly display the vast difference in speedup factors.}
    \label{fig:speedup_vs_seq}
\end{figure}

Figure~\ref{fig:speedup_vs_seq} presents the speedup achieved by the parallel CPU and parallel GPU implementations relative to the sequential CPU baseline. This bar chart clearly illustrates the substantial performance gains offered by both parallel approaches as the problem size increases, with the GPU demonstrating a significantly higher rate of improvement.

\begin{figure}[!ht]
    \centering
    \includegraphics[width=\columnwidth]{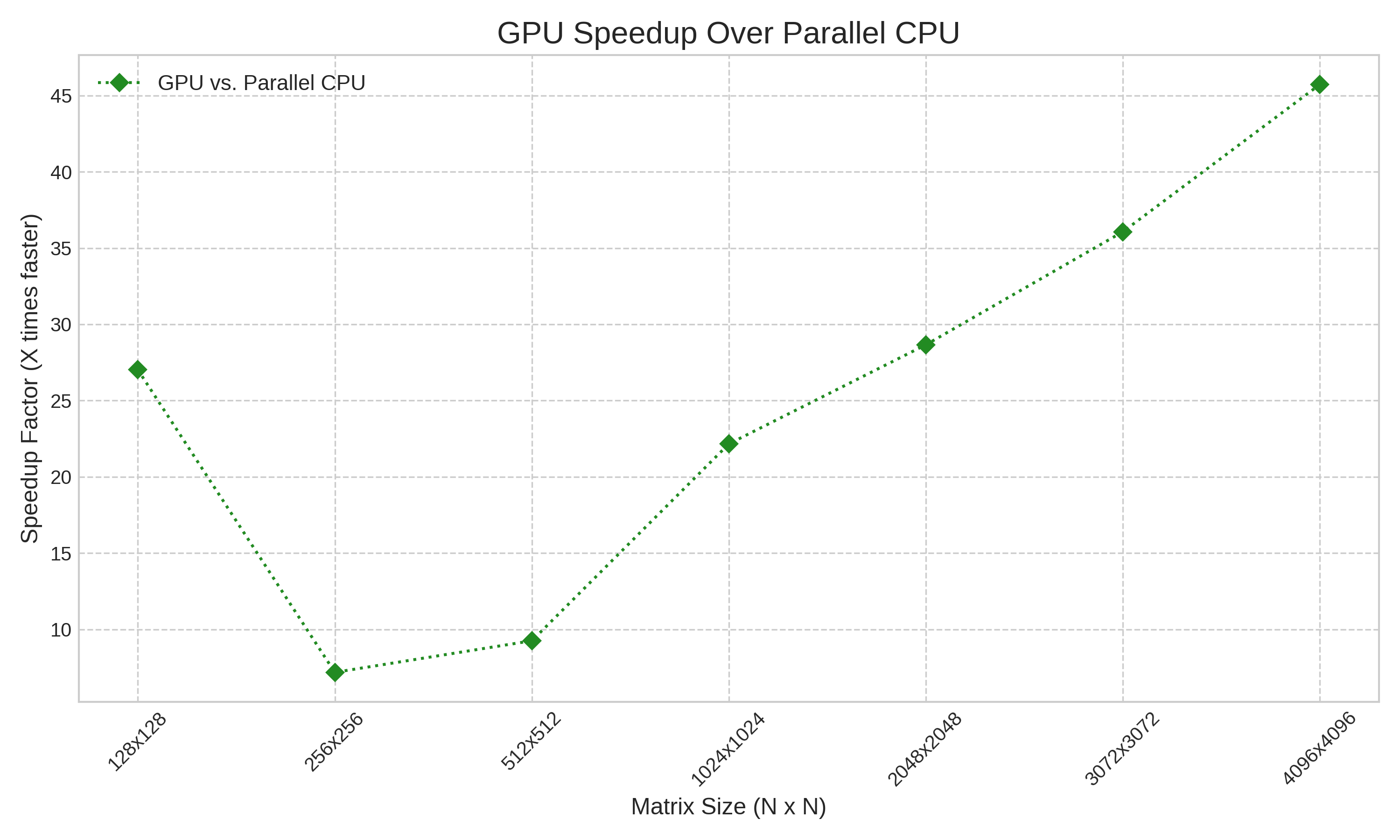}
    \caption{The performance advantage of the parallel GPU implementation relative to the parallel CPU implementation. The plot shows that the GPU's lead grows as the computational workload increases.}
    \label{fig:speedup_gpu_cpu}
\end{figure}

Finally, Figure~\ref{fig:speedup_gpu_cpu} provides a direct comparison of the two parallel architectures by plotting the speedup of the parallel GPU over the parallel CPU. This visualization isolates the performance advantage of the many-core GPU architecture relative to the multi-core CPU, showing a clear trend of increasing advantage as the matrix dimensions grow.

%======================================================================
\section{Discussion}
%======================================================================

The results presented in the previous section provide a clear quantitative measure of the performance differences between the multi-core CPU and many-core GPU architectures within a consumer-grade laptop. This section provides an interpretation of these results, contextualizes them within the broader academic literature, and acknowledges the scope and limitations of this study.

\subsection{Analysis of Results}

Our empirical data reveals three distinct performance regimes that highlight the fundamental architectural trade-offs between the AMD Ryzen 7 5800H CPU and the NVIDIA GeForce GTX 1650 Mobile GPU.

First, for small problem sizes, the cost of parallelization can outweigh its benefits. This phenomenon, known as parallel overhead, is evident in our results for the 128x128 matrix (Table~\ref{tab:results}), where the parallel OpenMP implementation is significantly slower than the sequential version (7.10\,ms vs. 2.18\,ms). This performance degradation is attributable to the overhead of thread creation, management, and synchronization on the CPU, which for a computationally trivial workload, consumes more time than is saved by parallel execution.

Second, as the matrix dimensions increase, the parallel CPU implementation demonstrates strong and consistent scaling. For matrices of size 256x256 and larger, the OpenMP version achieves a stable speedup of approximately 12-14x over the sequential baseline. This indicates that for computationally significant workloads, the OpenMP model effectively utilizes the 8 cores and 16 threads of the Ryzen 7 processor, providing a substantial and predictable performance improvement characteristic of mature shared-memory parallel programming.

Third, the results clearly demonstrate the asymptotic dominance of the GPU for large-scale, data-parallel tasks, even in a mobile form factor. While the parallel CPU provides a consistent speedup, the GTX 1650's performance advantage grows super-linearly as the matrix size increases. For the 4096x4096 matrix, the GPU is not only $\sim$593x faster than the sequential CPU but also over 45x faster than the highly-optimized parallel CPU implementation. This is a particularly striking result, showing that even a mobile, entry-level discrete GPU can dramatically outperform a powerful host CPU on a data-parallel task. This performance gain is a direct consequence of the GPU's many-core architecture, which exploits the immense parallelism inherent in matrix multiplication to a degree that a multi-core CPU, with its fewer but more complex cores, cannot match.

\subsection{Contextualizing with Existing Literature}

Our findings are in strong agreement with the conclusions of Lee et al.~\cite{lee2010debunking}, who argued that while GPUs offer a distinct performance advantage, claims of 100--1000x speedups often result from comparing an optimized GPU implementation against unoptimized sequential CPU code. Our results affirm this perspective; the speedup of the GPU is a formidable $\sim$45x when compared to an \textit{optimized, parallel} CPU implementation on the same machine. This figure is both impressive and realistic, aligning with their finding that the true performance gap between fully-optimized implementations on both architectures is significant but not mythological. Our work thus serves as a contemporary validation of this principle on a foundational HPC kernel, demonstrating its applicability even on consumer-grade heterogeneous hardware.

\subsection{Limitations of the Study}

It is important to acknowledge the boundaries of this research. First, the implementations, while robust, represent one approach among many. The CUDA kernel, though effective in its use of shared memory, could be further optimized with more advanced techniques such as register tiling or dynamic block sizing, which might yield additional performance. Second, this study was conducted on a single hardware configuration. The precise speedup ratios and performance crossover points are dependent on the specific CPU and GPU models used. We expect that higher-end hardware would yield larger absolute performance numbers, but the overall qualitative trends and architectural insights are likely to remain consistent.

%======================================================================
\section{Conclusion and Future Work}
%======================================================================

\subsection{Summary of Findings}

This study conducted an empirical performance comparison of dense matrix multiplication on a consumer-grade laptop equipped with an 8-core AMD Ryzen 7 CPU and an NVIDIA GeForce GTX 1650 Mobile GPU. Our results reaffirm that for computationally intensive, data-parallel tasks, the choice of architecture has a profound impact on performance, even on widely accessible hardware. We demonstrated that while an optimized, multi-threaded CPU implementation provides a substantial and consistent speedup over sequential execution, the massively parallel architecture of the mobile GPU offers a performance advantage that scales dramatically with the problem size. For large-scale matrices, the GPU implementation was orders of magnitude faster than the sequential baseline and significantly outperformed the parallel CPU version by a factor of over 45x. These findings underscore that the immense throughput capability of many-core architectures is not confined to high-end datacenter hardware, but is a critical and accessible tool for accelerating demanding computational kernels on modern heterogeneous systems.

\subsection{Future Work}

The results of this study suggest several promising directions for future research.

First, the GPU implementation could be enhanced by incorporating more advanced optimization techniques. While our current kernel effectively uses shared memory to reduce global memory traffic, further performance gains could be realized by implementing sophisticated memory tiling and register blocking strategies \cite{ryoo2008optimization}. Such methods could further improve data locality and reduce the latency of memory operations, potentially pushing performance closer to the hardware's theoretical peak.

Second, the comparative methodology employed in this paper could be extended to other important HPC kernels. Matrix multiplication is just one of the foundational "dwarfs" of scientific computing \cite{asanovic2006landscape}. A similar analysis of other common kernels, such as Fast Fourier Transforms (FFTs), stencil computations, or sparse matrix operations, would provide a more comprehensive understanding of the performance trade-offs between CPU and GPU architectures across a wider range of computational patterns. This would contribute valuable data to the broader discussion on workload characterization for heterogeneous systems.

Finally, a crucial next step would be to apply the optimized GPU kernel to a real-world scientific application to measure its end-to-end impact. While micro-benchmarks are essential for architectural analysis, integrating the accelerated kernel into a complete application---such as a deep learning training framework or a computational fluid dynamics simulation---would quantify the practical, wall-clock performance improvement on a complete scientific workflow. This would validate the real-world utility of the architectural advantages demonstrated in this paper.

% --- Bibliography ---
\bibliographystyle{IEEEtran}
\bibliography{main} % Assumes your file is named biblo.bib
\appendix
\section{Benchmark Source Code}
\label{sec:appendix_code}

The complete C++/CUDA source code used to generate all performance data in this paper is provided below for reproducibility.

\begin{lstlisting}[caption={Complete C++/CUDA source code for the benchmark.}, label={lst:benchmark_code}]
#include <iostream>
#include <vector>
#include <string>
#include <iomanip>
#include <cstdlib>
#include <ctime>
#include <chrono>
#include <omp.h>
#include <cuda_runtime.h>

// =================================================================
// CUDA Error Checking Wrapper
// =================================================================
#define CUDA_CHECK(err) { \
    cudaError_t err_ = (err); \
    if (err_ != cudaSuccess) { \
        std::cerr << "CUDA Error in " << __FILE__ << " line " << __LINE__ \
                  << ": " << cudaGetErrorString(err_) << std::endl; \
        exit(EXIT_FAILURE); \
    } \
}

// =================================================================
// Matrix Multiplication Implementations
// =================================================================

void initializeMatrix(std::vector<float>& matrix, int size) {
    for (int i = 0; i < size * size; ++i) {
        matrix[i] = static_cast<float>(rand()) / static_cast<float>(RAND_MAX);
    }
}

void sequentialMatrixMultiply(const std::vector<float>& A, const std::vector<float>& B, std::vector<float>& C, int size) {
    for (int row = 0; row < size; ++row) {
        for (int col = 0; col < size; ++col) {
            float sum = 0.0f;
            for (int k = 0; k < size; ++k) {
                sum += A[row * size + k] * B[k * size + col];
            }
            C[row * size + col] = sum;
        }
    }
}

void openmpMatrixMultiply(const std::vector<float>& A, const std::vector<float>& B, std::vector<float>& C, int size) {
    #pragma omp parallel for collapse(2)
    for (int row = 0; row < size; ++row) {
        for (int col = 0; col < size; ++col) {
            float sum = 0.0f;
            for (int k = 0; k < size; ++k) {
                sum += A[row * size + k] * B[k * size + col];
            }
            C[row * size + col] = sum;
        }
    }
}

__global__ void matrixMulGpu(const float* A, const float* B, float* C, int size) {
    int row = blockIdx.y * blockDim.y + threadIdx.y;
    int col = blockIdx.x * blockDim.x + threadIdx.x;

    if (row < size && col < size) {
        float sum = 0.0f;
        for (int k = 0; k < size; ++k) {
            sum += A[row * size + k] * B[k * size + col];
        }
        C[row * size + col] = sum;
    }
}

// =================================================================
// Main Benchmark Runner
// =================================================================
int main() {
    srand(static_cast<unsigned>(time(0)));

    std::vector<int> matrix_sizes = {128, 256, 512, 1024, 2048, 3072, 4096};

    // Print the header for our CSV data table
    std::cout << "Matrix_Size,Sequential_CPU_ms,Parallel_CPU_ms,Parallel_GPU_ms,"
              << "Speedup_CPU_vs_Seq,Speedup_GPU_vs_CPU,Speedup_GPU_vs_Seq" << std::endl;

    for (int N : matrix_sizes) {
        // --- Host Memory Allocation & Initialization ---
        std::vector<float> h_A(N * N);
        std::vector<float> h_B(N * N);
        std::vector<float> h_C(N * N, 0.0f); // Re-use one C matrix for all results

        initializeMatrix(h_A, N);
        initializeMatrix(h_B, N);

        // --- 1. Sequential CPU Benchmark ---
        auto start_seq = std::chrono::high_resolution_clock::now();
        sequentialMatrixMultiply(h_A, h_B, h_C, N);
        auto end_seq = std::chrono::high_resolution_clock::now();
        std::chrono::duration<double, std::milli> duration_seq = end_seq - start_seq;

        // --- 2. Parallel CPU (OpenMP) Benchmark ---
        auto start_omp = std::chrono::high_resolution_clock::now();
        openmpMatrixMultiply(h_A, h_B, h_C, N);
        auto end_omp = std::chrono::high_resolution_clock::now();
        std::chrono::duration<double, std::milli> duration_omp = end_omp - start_omp;

        // --- 3. Parallel GPU (CUDA) Benchmark ---
        float *d_A, *d_B, *d_C;
        size_t matrix_bytes = N * N * sizeof(float);
        CUDA_CHECK(cudaMalloc(&d_A, matrix_bytes));
        CUDA_CHECK(cudaMalloc(&d_B, matrix_bytes));
        CUDA_CHECK(cudaMalloc(&d_C, matrix_bytes));

        auto start_gpu = std::chrono::high_resolution_clock::now();
        CUDA_CHECK(cudaMemcpy(d_A, h_A.data(), matrix_bytes, cudaMemcpyHostToDevice));
        CUDA_CHECK(cudaMemcpy(d_B, h_B.data(), matrix_bytes, cudaMemcpyHostToDevice));
        dim3 threadsPerBlock(16, 16);
        dim3 numBlocks((N + threadsPerBlock.x - 1) / threadsPerBlock.x, (N + threadsPerBlock.y - 1) / threadsPerBlock.y);
        matrixMulGpu<<<numBlocks, threadsPerBlock>>>(d_A, d_B, d_C, N);
        CUDA_CHECK(cudaMemcpy(h_C.data(), d_C, matrix_bytes, cudaMemcpyDeviceToHost));
        auto end_gpu = std::chrono::high_resolution_clock::now();
        std::chrono::duration<double, std::milli> duration_gpu = end_gpu - start_gpu;

        CUDA_CHECK(cudaFree(d_A));
        CUDA_CHECK(cudaFree(d_B));
        CUDA_CHECK(cudaFree(d_C));
        
        // --- 4. Calculate Speedups ---
        double speedup_cpu_vs_seq = duration_seq.count() / duration_omp.count();
        double speedup_gpu_vs_cpu = duration_omp.count() / duration_gpu.count();
        double speedup_gpu_vs_seq = duration_seq.count() / duration_gpu.count();

        // --- 5. Report Results for this Size ---
        std::cout << N << "x" << N << ","
                  << std::fixed << std::setprecision(4) << duration_seq.count() << ","
                  << std::fixed << std::setprecision(4) << duration_omp.count() << ","
                  << std::fixed << std::setprecision(4) << duration_gpu.count() << ","
                  << std::fixed << std::setprecision(2) << speedup_cpu_vs_seq << "x,"
                  << std::fixed << std::setprecision(2) << speedup_gpu_vs_cpu << "x,"
                  << std::fixed << std::setprecision(2) << speedup_gpu_vs_seq << "x" << std::endl;
    }

    std::cout << "\nBenchmark complete." << std::endl;

    return 0;
}
\end{lstlisting}

\end{document}